\documentclass[usenatbib]{mn2e}
\usepackage{graphicx}

\title[$L_X$--SFR relation in star forming galaxies]
{${\bmath L_{\bmath X}}$--SFR relation in star forming galaxies} 
 
\author[M.Gilfanov et al.]
       {M.~Gilfanov$^{1,2}$, H.-J. Grimm$^1$, R.~Sunyaev$^{1,2}$\\
        $^1$Max-Planck-Institut f\"ur Astrophysik, 85741 Garching
            b. M\"unchen, Germany\\
	$^2$Space Research Institute, Moscow, Russia}
\date{\today}

\pubyear{2002}

\begin{document}

\maketitle

\begin{abstract}
We compare the results of \citet{grimm} and \citet{ranalli} on
the $L_X$--SFR relation in normal galaxies. Based on the $L_X-$stellar mass 
dependence for LMXBs, we show, that low SFR 
($\la 1$ M$_{\sun}$/year) galaxies in the Ranalli et al. sample are
contaminated by the X-ray emission from low mass
X-ray binaries, unrelated to the current star formation activity.

The most important conclusion from our comparison is, however, that 
after the data are corrected for the ``LMXB contamination'', the two
datasets become consistent with each other, despite of their
different content, variability effects, difference in the adopted
source distances, X-ray flux  and star formation  rate determination
and  in the cosmological parameters used in  interpreting the HDF-N
data. They also agree well, both in the low and high SFR regimes,
with the predicted $L_X$--SFR dependence derived from the parameters
of the ``universal'' HMXB luminosity function.  This encouraging result 
emphasizes the potential of the X-ray luminosity as an independent 
star formation rate indicator for normal galaxies. 

\end{abstract}

\section{Introduction}
\label{sec:intro}

\begin{figure}
\resizebox{\hsize}{!}{\includegraphics{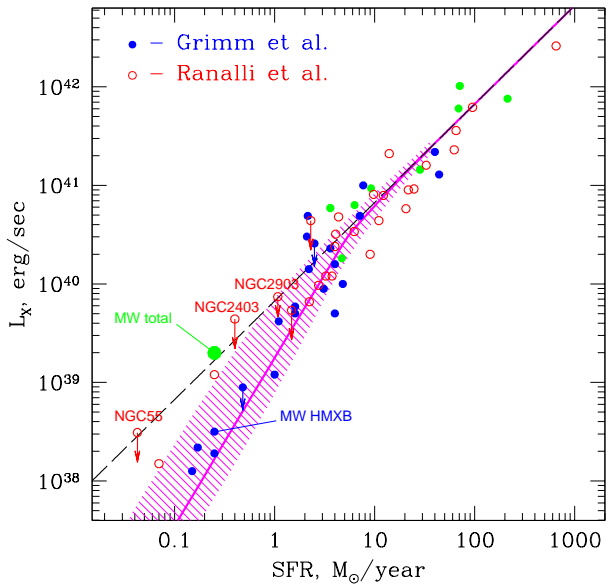}}
\caption{$L_X$--SFR relation. All points from \citet{ranalli} and
\citet{grimm}. The galaxies with the expected LMXB fraction exceeding
50\% are plotted as upper limits.  
The thick solid line shows predicted relation between the most probable
value of $L_X$ and SFR, the shaded area -- it's 67\% intrinsic spread. 
The straight dashed line shows the expectation mean for $L_X$, which would
be obtained if X-ray luminosities of  many galaxies with similar
SFR were averaged. To demonstrate importance of the LMXB  
contribution at low SFR/M$_*$, both HMXB and total
luminosities are plotted for the Milky Way.
}
\label{fig:all_points}
\end{figure}

Based on Chandra observations of nearby star forming galaxies
and studies of high mass X-ray binaries (HMXB) population in the
Milky Way and SMC, \citet*{grimm} proposed recently that high mass
X-ray binaries can be used as a star formation rate (SFR) indicator. 
They found, that in a broad  range of star formation regimes and rates
the X-ray luminosity distribution of HMXBs can be approximately
described by a ``universal'' luminosity function -- 
a power law with the slope of $\sim 1.6$ and a cut-off at
$\lg(L_X)\sim 40.5$, which normalization is proportional to the SFR. 
As the 2--10 keV luminosity  $L_X$  of a normal galaxy with
sufficiently high $SFR/M_*$ ratio ($M_*$ -- total stellar mass)  is
dominated by the emission from high mass X-ray binaries,
the X-ray luminosity can be used as a star formation rate indicator
for normal galaxies.

Although the normalization of the luminosity function and the number
of sources are proportional 
to the SFR, the $L_X$--SFR dependence is non-linear in the low SFR
regime and becomes linear only at sufficiently high values of SFR
(thick solid line in Fig.\ref{fig:all_points}). This non-linear
behavior at low SFR values is {\bf not} related to intrinsic
non-linear SFR dependent effects in the population of the HMXB sources.
It is rather caused by the fact that the quantity of
interest is a sum of the luminosities of discrete sources  -- 
$L_{X,\rm tot}=\sum_k L_{X,k}$, with $L_{X,k}$ obeying a power law 
luminosity distribution.  The non-linear behavior is caused by the
properties of $p(L_{X,\rm tot})$ probability distribution, namely, the  
difference between its expectation mean (average) and its
mode (most probable value). 
This effect was discussed in \citet{grimm} and \citet{lmxb} and will
be given a detailed treatment in \citet*{stat}. 
 The position of the break in the $L_X$--SFR relation 
is defined by the parameters of the luminosity function. For
particular values of the slope and cut-off luminosity found by
\citet{grimm}, the boundary between non-linear and linear regime lies
at SFR$\sim 4.5$ M$_{\sun}$/year or, equivalently, $L_X\sim 3\cdot
10^{40}$ erg/sec. Chandra and ASCA measurements of the
total X-ray luminosity of a number of nearby star forming
galaxies were in a good qualitative and quantitative agreement with the
predicted  $L_X$--SFR relation (Fig.\ref{fig:all_points}, thick solid
curve and filled circles). Moreover, the distant star forming galaxies,
observed by Chandra in the Hubble Deep Field North \citep{brandt01} 
at redshifts of $z\sim 0.2-1.3$, also obey the same relation. In 
the linear high SFR regime it is given by:
\begin{equation}
SFR[{\rm M_{\sun}/yr}]=\frac{L_{\rm 2-10~keV}}{6.7\cdot 10^{39}~ \rm erg/s}
\label{eq:grimm}
\end{equation}
where SFR is the formation rate of massive stars, $M>5~M_{\odot}$.
\citet{grimm} pointed out importance of two contaminating factors,
unrelated to the current star formation activity:
(i) emission of the central supermassive black hole, which even in the
low luminosity AGNs can easily out-shine X-ray binaries and 
(ii) contribution of the low mass X-ray
binaries, which might be especially important in the low SFR regime.

\citet*{ranalli} independently studied X-ray luminosity of normal
galaxies using 
the ASCA and BeppoSAX archival data and Chandra observations of the
HDF-N and found a tight correlation
between their X-ray, radio (1.4 GHz) and FIR fluxes. They suggested
that the 2--10 keV luminosity of normal galaxies can be used as a
SFR indicator and derived the  relation:
\begin{equation}
SFR[{\rm M_{\sun}/yr}]=\frac{L_{\rm 2-10~keV}}{5\cdot 10^{39}~ \rm erg/s}
\label{eq:ranalli}
\end{equation}
This  formula agrees reasonably well with that obtained by
\citet{grimm} for the high SFR regime, eq.(\ref{eq:grimm}).
However, \citet{ranalli} noted, that the $L_X$--SFR relation was
linear in the entire range of the star formation rates, including the
low SFR regime, in apparent contradiction to \citet{grimm} results. 

In this Letter we compare  the \citet{grimm} and \citet{ranalli}
samples of the galaxies. We demonstrate, that the X-ray emission from
the low SFR galaxies in the  \citet{ranalli} sample is likely to be
``contaminated'' by  low mass X-ray binaries, which
are unrelated to current star formation activity. 
After the ``LMXB contamination'' is accounted for, the two datasets 
agree qualitatively and  quantitatively and are  consistent with the 
$L_X$--SFR relation expected on the basis of the  ``universal'' HMXB
luminosity function derived by \citet{grimm}.

\begin{figure}
\hbox{
\resizebox{\hsize}{!}{\includegraphics{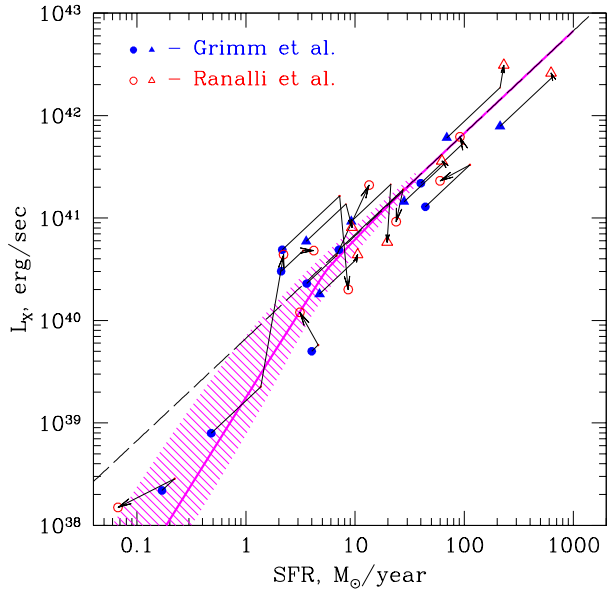}}
}
\caption{Comparison of the data for local (circles) and HDF-N (triangles)
galaxies present both in \citet{grimm} and \citet{ranalli}
samples. For each galaxy, it's positions in two samples
are connected by a  broken line with the arrow
directed from G to R. The first segment of each broken line shows the effect
of the difference in the source distance or cosmological parameters, 
the second segment shows cumulative effect of other factors, such as
variability and difference in the SFR values.  
}
\label{fig:compare}
\end{figure}

\section{The samples}

In the following we denote \citet{ranalli} and \citet{grimm} samples
as R and G correspondingly. The data from both samples are plotted
together in Fig.\ref{fig:all_points}.

\subsection{The local galaxies}

The two samples, although differently constructed, overlap
substantially, with 9 galaxies (out of 23 in each sample),
present in both. The sample R was derived using more rigorously
defined construction algorithm.
In almost all cases the authors adopted different distances and
different values of SFR. \citet{grimm} derived SFR values  averaging
the results of several independent  
estimators based on UV, FIR, H$_\alpha$ and radio flux measurements,
whereas \citet{ranalli} used radio flux measurements.  
The X-ray fluxes were obtained from different observations, sometimes
by different instruments and are, obviously, affected by variability
of the X-ray emission from the 
galaxies. For some of the galaxies the X-ray luminosity was
calculated by \citet{grimm} as a direct sum of the luminosities of
compact sources detected by Chandra. 

The Fig.\ref{fig:compare} compares positions of the
galaxies present in the both samples in the $L_X$--SFR plane.   
Note, that the difference in the  adopted distances does not have
effect at high values of SFR where the $L_X$--SFR relation is linear,
but it might destroy the correlation in the non-linear low SFR
regime.

\subsection{Hubble Deep Field North}

Both \citet{grimm} and \citet{ranalli} used similar selection criteria. 
Each sample contains seven sources, of which six are present in both
samples. The sources \#185 and \#148 (according to Table 2 in
\citet{brandt01})  are absent from the R and G
correspondingly.  The latter was excluded from sample G because no 
1.4 GHz flux was detected, with the upper limit of 23
$\mu$Jy \citep{richards98}.
The main difference lies in computing the X-ray fluxes and
luminosities. \citet{grimm} used 2--8 keV fluxes from Chandra catalog
and K-corrected them to 2--10 keV rest-frame luminosity using the
spectral indexes from \citet{brandt01}. \citet{ranalli}
derived the X-ray count rates in two redshift-corrected energy bands
and based their final K-correction on the 
recomputed spectral indexes. The following cosmological parameters
were used: $H_0=50$ km/s/Mpc, $q_0=0.1$ (sample R) and  $H_0=70$
km/s/Mpc, $q_0=0.5$, $\Lambda=0$ (sample G). 
The positions of the data points in the $L_X$--SFR plane are compared in 
Fig.\ref{fig:compare}.

\section{LMXB contribution}

\begin{figure}
\resizebox{\hsize}{!}{\includegraphics{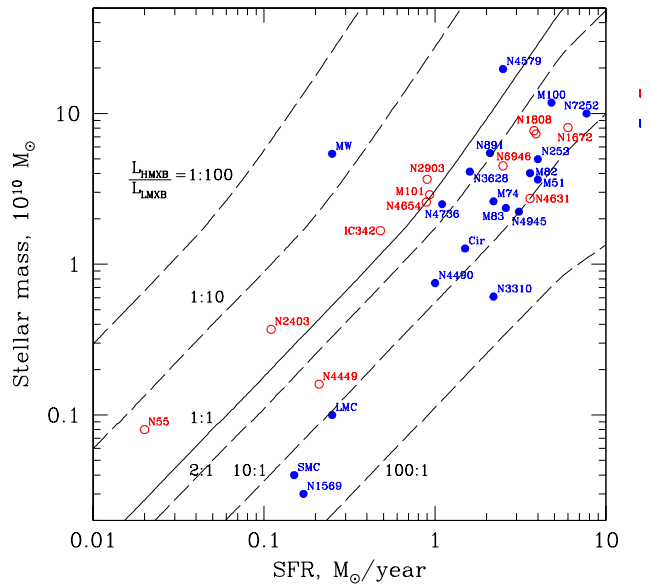}}
\caption{Location of galaxies from R (open circles) and G (filled
circles) samples on the SFR$-M_*$ plane. The dashed lines correspond
to constant ratio of the the most probable values of HMXB and LMXB
luminosities estimated from respective average luminosity functions
with account for non-linear effects of statistics. For the points
above the solid line, the LMXB  contribution exceeds 50\%.
}
\label{fig:sfr_mass}
\end{figure}

Due to long evolutionary time scale, the population of low mass X-ray
binaries is unrelated to the current star formation activity. It is,
rather, proportional to the stellar mass of the host galaxy
\citep{lmxb}. Hence,  the X-ray  emission from LMXBs can contaminate 
the $L_X$--SFR relation, as exemplified by the Milky Way galaxy, in
which the  LMXBs contribution exceeds $\approx 90\%$
\citep*[Fig.\ref{fig:all_points},][]{grimm1}.    
Although  LMXB and HMXB sources can not be easily separated 
based on the X-ray data, and optical identifications
are (potentially) available only for the most nearby galaxies, 
the number and combined luminosity of LMXBs can be sufficiently
accurately predicted based on the stellar mass of the host galaxy
\citep{lmxb}. Thus, relative contributions of LMXB and HMXB sources to
the X-ray luminosity of the galaxy are defined by its position on the 
SFR$-M_*$ plane (Fig.\ref{fig:sfr_mass}).

Stellar masses of the galaxies were calculated using K-band
magnitudes from 2MASS Large Galaxy Atlas \citep{2mass_lga} with
the color based correction to the mass-to-light
ratio \citep{m2l}. The mass of the Milky Way was calculated using its
K-band luminosity  obtained by \citet{malhotra96} from 3D modeling of
the DIRBE data, and assuming the same mass-to-light ratio as in M31. 
The stellar masses of LMC and SMC were estimated from their dynamical
masses \citep{grimm}, assuming $M_{\rm dyn}/M_*=5$.
The distances for the galaxies from the sample G
are same as in \citet{grimm}. We re-examined the distance to low SFR
galaxies. For NGC55 (1.6 Mpc) and M101 (7.2 Mpc) we adopted values
from \citet*{ngc55} and \citet*{m101}.  
The distances to NGC2403 (3.7 Mpc), NGC2903 (9.5 Mpc), NGC4449 (3.8
Mpc) and NGC4654 (17.6 Mpc) were estimated from IR Tully-Fisher
relation \citep{a82} using data from \citet{tb95} and calibration from 
\citet{sakai00}.  The distances to other galaxies from R sample are
the same as in \citet{ranalli}. 

The galaxies from R and G samples are plotted in the SFR$-M_*$ plane
in Fig.\ref{fig:sfr_mass}, along with the contours of constant $L_{\rm
HMXB}:L_{\rm LMXB}$ ratio. The luminosities of LMXB and HMXB sources 
were estimated from their respective average luminosity functions
obtained by \citet{grimm} and \citet{lmxb}. In estimating the LMXB
luminosity we used average normalization for late 
type galaxies.  Although in the limit of large number of
sources linear relations hold, $L_{\rm LMXB}\propto M_*$ and
$L_{\rm HMXB}\propto {\rm SFR}$, the contours are not
straight  lines at M$_*\la 2\cdot 10^{10}$ M$_{\odot}$ and SFR$\la 4$
M$_{\odot}$/yr due to effects of statistics \citep{stat}.

As expected, the ``LMXB contamination'' plays role mostly at low SFR
values and becomes unimportant at high star formation rates.
In all but one galaxies from the sample R having SFR$\la 1$
M$_{\sun}$/yr, the expected contribution of LMXBs exceeds $\sim
50\%$. These galaxies are shown in Fig.\ref{fig:all_points} as upper
limits. For two galaxies (NGC55 and NGC2403) mostly deviating from the
common trend in Fig.\ref{fig:all_points}, the expected LMXB
contribution exceeds $\sim 70\%$.

\begin{figure}
\resizebox{\hsize}{!}{\includegraphics{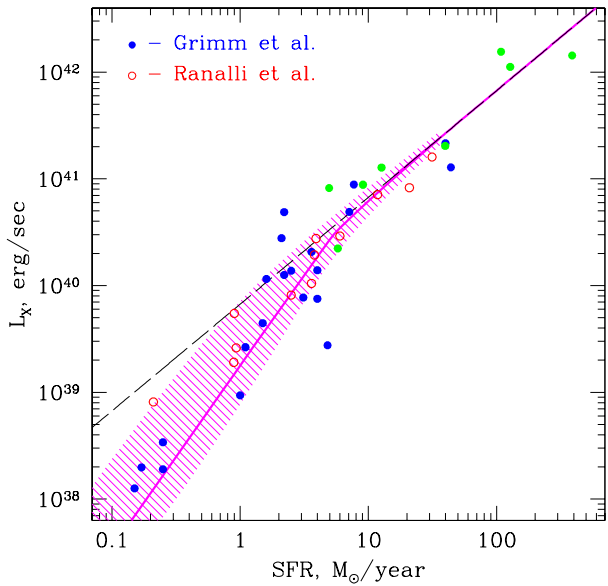}}
\caption{$L_X$--SFR relation -- combined data from \citet{ranalli} and
\citet{grimm}, with duplications excluded. The $L_X$ values for nearby
galaxies were corrected for the LMXB contribution estimated from the
stellar mass. Three galaxies with more than 50\% LMXB contribution and
small stellar mass, $M_*<2\cdot 10^{10}$ M$_{\odot}$, for which large
intrinsic dispersion of the $L_X-M_*$ relation precludes accurate
estimate of the LMXB luminosity, are not plotted. 
The luminosities for the HDFN and Lynx field galaxies were computed
for $H_0=70$ km/s/Mpc, $\Omega_m=0.3$, $\Lambda=0.7$.  
The solid and dashed lines and the shaded area are the same as in
Fig.\ref{fig:all_points}. 
}
\label{fig:combined}
\end{figure}

\section{Combined sample and predicted ${\bmath L_{\bmath X}}$--SFR
relation} 

The Fig.\ref{fig:combined} shows all data from the samples G and R, excluding 
duplications. The tightness of $L_X-M_*$ relation at large $M_*$
(Fig.14 in \citealt{lmxb}) allows one to approximately correct
observed luminosities for the LMXB contribution. This correction was
applied for all galaxies, except those with $M_*< 2\cdot 10^{10}$
M$_{\odot}$ and   $L_{\rm HMXB}:L_{\rm LMXB}<1$. 
The latter (NGC55, NGC2403 and IC342) are not
plotted in  Fig.\ref{fig:combined}.

The solid curve in Fig.\ref{fig:combined} shows the predicted
$L_X$--SFR relation, calculated using the
parameters of the ``universal'' HMXB luminosity function derived by
\citet{grimm} from analysis of  five nearby star forming
galaxies with best known luminosity functions.    
It corresponds to  the mode of the probability distribution  -- the
{\em most likely} value of the X-ray luminosity of a randomly
chosen galaxy.  
The dashed line, on the contrary, shows the expectation mean -- the
value, that would result 
from {\em averaging} of the X-ray luminosities of many galaxies
having similar values of SFR. Due to the properties of the probability
distribution of the total luminosity of a population of discrete
sources, $L_{X,\rm tot}=\sum_k L_{X,k}$,
these two quantities are not identical in the low SFR limit, when the
number of sources is small.

Due to skewness of the probability distribution $p (L_{X,\rm tot})$
(Fig.2 in \citealt{lmxb}), large and asymmetric dispersion around the
solid curve in Fig.\ref{fig:combined} is expected in the non-linear
low SFR regime. The probability to find a 
galaxy   below the curve is $\approx 12-16\%$ at SFR=$0.2-1.5~M_{\odot}$/yr
and increases to $\approx 30\%$ at SFR=$4-5 ~M_{\odot}$/yr, near the break
of the $L_X$--SFR relation.
Of course in the linear regime (SFR$\ga 10~M_{\odot}$/yr) it
asymptotically approaches $\sim 50\%$, as expected. 
This asymmetry is already seen from the distribution of the points  in
Fig.\ref{fig:combined} -- at low SFR values there are more points
above the solid curve, than below. 
Moreover, the low probability high luminosity tail of the  
$p (L_{X,\rm tot})$ distribution will lead to appearance of
galaxies-outliers with significantly larger than expected value of the
total luminosity. Such galaxies will inevitably appear as the plot is
populated with more objects.  Non-gaussianity of the $p (L_{X,\rm
tot})$ distribution makes  least square and $\chi^2$ fitting
techniques inadequate for  analysis of the $L_X$--SFR
relation in the low SFR regime.

\section{Conclusion}

We compared results of \citet{grimm} and \citet{ranalli} on
relation of the X-ray luminosity and the star formation rate in normal 
galaxies (Fig.\ref{fig:all_points} and \ref{fig:compare}). 

Addressing the discrepancy in the low SFR regime, we note that six out
of seven galaxies from \citet{ranalli}, having SFR$\la 1$ M$_{\sun}$/yr,
are likely to be contaminated by the X-ray emission from low mass
X-ray binaries, having no relation to the  current star formation
activity. 
Furthermore, at $M_*\la 10^{10}~M_{\odot}$ and SFR$\la 1$
M$_{\odot}$/yr, the expected luminosity of X-ray binaries does not
exceed $\la 10^{39}$ erg/s. This is comparable or smaller than that of
low luminosity AGNs often found by Chandra in otherwise apparently normal
galaxies. The AGN contribution can not be identified and separated,
unless high angular resolution imaging data are available.
Secondly, the probability distribution of the total
luminosity of a population of discrete sources, 
$L_{X,\rm tot}=\sum_k L_{X,k}$, is significantly non-Gaussian for low
values of $L_{X,\rm tot}$.
This should not be ignored when analyzing and interpreting the
$L_X$--SFR relation in the low SFR regime.  

The most  important conclusion is, however, that  
after the potentially ``LMXB contaminated'' galaxies are excluded, the
two datasets {\em become consistent} with each other, 
despite of their different content, variability effects,
difference in the adopted source distances, X-ray flux 
and star formation  rate determination and  in the cosmological
parameters used in  interpreting the HDF-N data. 
The $\sim 30\%$ difference in the calibration
of the $L_X$--SFR relation is insignificant considering the number and
amplitude of the uncertainties involved. 
They also agree well, both in the low and high SFR regimes,  with the
predicted $L_X$--SFR dependence derived from the parameters of the
``universal'' HMXB luminosity function (Fig.\ref{fig:combined}). 
This is an encouraging result emphasizing the potential of X-ray luminosity
as an independent star formation rate indicator.  

\section{Acknowledgements}

We are grateful the referee, Dr. Pranab Ghosh, for critical and
stimulating comments on the original manuscript which helped to
improve the paper.  

{}


\begin{thebibliography}{}

\bibitem[\protect\citeauthoryear{Aaronson et al.}{1982}]{a82}
Aaronson M. et al., 1982, ApJS, 50, 241

\bibitem[\protect\citeauthoryear{Bell \& de Jong}{2001}]{m2l}
Bell E. \& de Jong R., 2001, ApJ, 550, 212

\bibitem[\protect\citeauthoryear{Brandt et al.}{2001}]{brandt01}
Brandt W.N. et al., 2001, AJ, 122, 2810

\bibitem[\protect\citeauthoryear{Gilfanov}{2003}]{lmxb} Gilfanov M.,
submitted to MNRAS, astro-ph/0309454

\bibitem[\protect\citeauthoryear{Gilfanov, Grimm \& Sunyaev}{Gilfanov
et al.}{2003}]{stat} Gilfanov M., Grimm H.-J. \& Sunyaev R., 2003, in
preparation 

\bibitem[\protect\citeauthoryear{Grimm, Gilfanov \& Sunyaev}{Grimm et
al.}{2002}]{grimm1} Grimm H.-J., Gilfanov M.\& Sunyaev R., 2002, A\&A,
391, 923 

\bibitem[\protect\citeauthoryear{Grimm, Gilfanov \& Sunyaev}{Grimm et
al.}{2003}]{grimm} Grimm H.-J., Gilfanov M.\& Sunyaev R., 2003, MNRAS,
339, 793 


\bibitem[\protect\citeauthoryear{Jurcevic, Pierce \& Jacoby}{Jurcevic et
al.}{2000}]{m101} Jurcevic J.S., Pierce M.J. \& Jacoby G.H., 2000, MNRAS,
313, 868

\bibitem[\protect\citeauthoryear{Jarrett, Chester \& Cutri}{Jarrett et
al.}{2000}]{2mass_lga} Jarrett T.H., Chester T., \& Cutri R., 2003,
AJ, 125, 525

\bibitem[\protect\citeauthoryear{Malhotra et al.}{Malhotra et al.}
{1996}]{malhotra96} Malhotra S. et al., 1996, ApJ, 473, 687

\bibitem[\protect\citeauthoryear{Puche, Carignan \& Wainscoat}{Puche et al.}
{1991}]{ngc55} Puche D., Carignan C. \& Wainscoat R., 1991, AJ, 101, 447

\bibitem[\protect\citeauthoryear{Ranalli, Comastri \& Seti}{Ranalli et
al.}{2003}]{ranalli} Ranalli P., Comastri A. \& Seti G., 2003, A\&A,
399, 39

\bibitem[\protect\citeauthoryear{Richards et al.}{1998}]{richards98}
Richards E.A. et al., 1998, Astron.J., 116, 1039

\bibitem[\protect\citeauthoryear{Sakai et al.}{Sakai et al.}
{2000}]{sakai00} Sakai S. et al., 2000, ApJ, 529, 698

\bibitem[\protect\citeauthoryear{Tormen \& Burstein}{Tormen \& Burstein}
{1995}]{tb95} Tormen G. \& Burstein D., 1995, ApJS, 96, 123


\end{thebibliography}
\end{document}